\documentclass[11pt]{article}
\usepackage{jcappub}

\usepackage{amsmath}
\usepackage{amssymb}
\usepackage{graphicx}
\usepackage{epsfig}
\usepackage{caption}
\usepackage{subcaption}
\begin{document}

\title{The TeV Blazar Measurement of the Extragalactic Background Light}

\author{Rebecca Reesman$^{1,2}$}
\emailAdd{rreesman@physics.osu.edu}
\author{T.P. Walker$^{1,2,3}$}
\affiliation{$^{1}$Department of Physics, The Ohio State University, Columbus, OH 43210}
\affiliation{$^{2}$Center for Cosmology and Astro-Particle Physics, The Ohio State University, Columbus, OH 43210} 
\affiliation{$^{3}$Department of Astronomy, The Ohio State University, Columbus, OH 43210}

\abstract{
Gamma-rays propagating through space are likely to be extinguished via electron-positron pair production off of the ambient extragalactic background light (EBL). The spectrum of the EBL is produced by starlight (and starlight reprocessed by dust) from all galaxies throughout the history of the Universe. The attenuation of 40 - 400 GeV gamma-rays has been observed by \textit{Fermi} and used to measure the EBL spectrum over energies 1 eV -10 eV out to redshift $z\sim 1$.   Measurements of several TeV blazers are consistent with attenuation, attributed to the EBL at redshift $z\sim 0.1$.  Here we simultaneously analyze a set of TeV blazers at $z\sim 0.1$ to measure the optical depth for 100 GeV - 10 TeV gamma-rays, which interact with EBL of energies 0.05 eV - 5 eV.  Using a suite of models for the EBL, we show that the optical depth indicated by TeV blazar attenuation is in good agreement with the optical depths measured by \textit{Fermi} at lower gamma-ray energies and higher redshifts.}


\maketitle 

\section{Introduction}
Very high energy (VHE) (100 GeV - 10 TeV) gamma-rays which travel from cosmic sources can pair produce off ambient background photons provided the ambient photon energy is greater than 5 eV  - $5 \times10^{-2}$ eV, respectively. The extragalactic background light (EBL) is composed of the total light emitted by stars over the history of the Universe. Its intensity shows two broad peaks at roughly 1eV (starlight) and at 0.01eV (starlight reemitted by  cold interstellar dust).  Therefore, starlight and starlight reprocessed by interstellear dust are expected to be the dominant contributors to the optical depth for these VHE gamma-rays. The energy density of the EBL is expected to be large enough to attenuate gamma-rays with energies $\sim$1 TeV ($\sim$ 100 GeV) which travel further than $z \sim 0.1$ ($z \sim 1$)\cite{nikishov, Jelley, Gould, Stecker:1992wi}. For current reviews see \cite{Costamante:2013sva, Dwek:2012nb}.

There are several approaches to calculating the EBL. Backward evolution models start with present day luminosity functions and trace them back in time over redshift \cite{Stecker:2005qs} . Forward evolution models  begin with initial cosmological conditions and propogate galaxy emissivities forward through time \cite{Primack:1998wn, Somerville:1998yg, Gilmore:2009zb}. These are also typically labelled semi-analytic models; they describe the cosmological evolution of dark matter and other structure formation in the Universe and  predict the EBL as a by-product. This is a more challenging approach, but it does offer extra insight into the astrophysical processes that contribute to the EBL. There are also methods that involve interpolation and extrapolation of observables over a range of redshifts \cite{Finke:2009xi, Dominguez:2010bv, Franceschini:2008tp}.

Directly measuring the EBL is complicated by high foreground contamination, such as zodiacal light, but there are some limiting bounds. Absolute intensity measurements are used to define an upper limit to the EBL at energies of $\sim$ 1 eV \cite{Dwek:2012nb}. Lower limits on the EBL intensity at $\sim$ few eV can be provided by deep galaxy counts \cite{Madau:1999yh, Helgason:2012xj}. In the lower panel of Fig. \ref{fig:tau} we show a representative model of the EBL as well as representative constraints from experiments, over the EBL energy range relevant to TeV gamma-rays.

The EBL can also be measured by the absorption of gamma-rays from distant sources. The Fermi Large Area Telescope (\textit{Fermi}-LAT) \cite{Atwood:2009ez} has measured the attenuation of gamma-rays having energies $\sim$ 40 GeV - 400 GeV (corresponding to EBL absorption by photons of energies in the range 1 eV - 10 eV). 
As we discuss in the conclusion, other studies have used a variety of techniques to constrain the EBL attenuation, using various proxies for the source's intrinsic TeV spectrum.  Here we use a statistical method which marginalizes over the source parameters and combines the measurements from a set of TeV sources clustered around $ z \sim 0.1$.

\textit{Fermi} \cite{Ackermann:2012sza} shows that the gamma-ray attenuation is consistent with EBL model predictions at $z \sim 1$.  We show that the combined TeV data sets of H.E.S.S., Veritas, and Magic, analyzed in a similar way, are consistent with the same model predictions for the energy range  $5\times10^{-2}$ eV - 5 eV and a mean redshift of $z\sim 0.1$.  In particular, they agree well over the range where they are probing the same EBL energies of  1 eV - 5 eV (i.e. for gamma-rays of energy 100 GeV -- 400 GeV).  At energies of a few eV, the EBL should be dominated by the intergrated light produced by all stars throughout the history of the Universe - agreement between the two data sets, taken by different techniques and sensitive to two different epochs of the Universe,  re-enforces our understanding of the star formation history.  Lastly, we present a complete measurement of the optical depth of the Universe for gamma-rays of energies 40 GeV -- 10 TeV.

\section{Method}
\subsection{Formalism}
As gamma-rays traverse intergalactic space they can interact with EBL photons if there is sufficient energy in the center of mass frame to create an electron-positron pair, namely twice the electron mass, 2m$_e$: 

\begin{equation}
\sqrt{2E_{\gamma} E_{EBL} (1-\cos\theta)} \geq 2 m_e,
\end{equation}
where E$_{\gamma}$ is the gamma-ray energy, E$_{EBL}$ is the EBL photon energy, and $\theta$ is the angle of incidence. In all equations we have set $c=1$.  The threshold for pair production is:
\begin{equation}
E_{EBL}^{thresh}=\frac{2m_e^2}{E_\gamma (1-\cos\theta)}
\end{equation}
The pair production cross-section is 

\begin{eqnarray}
\lefteqn{\sigma(E_1,E_2,\theta) = \frac{3\sigma_T}{16}(1-\beta^2)} \nonumber \\
& & \times \left[ 2\beta(\beta^2-2)+ (3-\beta^4)\ln \left( \frac{1+\beta}{1-\beta}\right)\right], 
\end{eqnarray}
where
\begin{equation}
\beta = \sqrt{1-\frac{2m_e^2}{E_{EBL} E_{\gamma} (1-\cos\theta)}},
\end{equation}
and $\sigma_T$ is the Thomson cross-section.

This process can attenuate the spectra of cosmic sources provided the EBL density is large enough and/or the source distance is sufficiently far. The optical depth relates the total distance a gamma-ray travels to its mean free path for interaction. As a full integral it is: 

\begin{eqnarray}
\lefteqn{\tau(E_\gamma,z_0) =  \frac{1}{2}\int^{z_0}_0 dz\;\frac{dl}{dz}\int^1_{-1}d(\cos\theta) \; (1-\cos\theta)} \nonumber \\ 
& & \times \int^{\infty}_{E_{min}} dE_{EBL}\; n(E_{EBL},z)\;\sigma(E_\gamma (1+z),E_{EBL},\theta).
\end{eqnarray}
where $dl/dz$ is the cosmological line element, defined as: $\frac{dl}{dz}=\frac{1}{(1+z)H_0} \frac{1}{\sqrt{\Omega_m(1+z)^3+\Omega_\Lambda}}$.

\subsection{Data Set}
 We chose to focus our search around $z\sim 0.1$ (mean: 0.104 $\pm$ 0.031, median: 0.102) so as to make a straight forward determination of the optical depth, which depends on redshift (the few TeV sources at higher redshift $z\sim 0.4$, PG 1553+113 and 3C 66A,  yield optical depths consistent with our analysis).  Our sample includes 7 BL Lacertae blazars detected by various Cherenkov instruments as VHE point sources. Details of the sources can be found in Table \ref{table:sources}. 

\begin{table}
\begin{center} 
\begin{tabular}{|ccccc|}
\hline
Source Name & z  & $\Gamma_{VHE}$ & Detector & Reference \\ \hline
BL Lac & 0.069 & 3.64 $\pm$ 0.54 & Magic & \cite{Albert:2007iv} \\
IES 0229+200 & 0.140 & 2.50 $\pm$ 0.19 $\pm$ 0.10 & HESS & \cite{Aharonian:2007wc} \\
IES 0806+524 & 0.138 & 3.6 $\pm$ 1.0 $\pm$ 0.3 & VERITAS & \cite{Acciari:2008cn} \\
J0152+017 & 0.080 & 2.95$\pm$0.36$\pm$0.20 & HESS & \cite{Aharonian:2008xv} \\
J0710+591 & 0.125 & 2.69$\pm$0.26$\pm$0.20 &  VERITAS & \cite{Acciari:2010qw} \\
PKS 2005-489 & 0.071 & 4.0$\pm$0.4 & HESS & \cite{Aharonian:2005cu} \\
W Comae & 0.102 &  3.81$\pm$0.35$\pm$0.34 & VERITAS & \cite{Acciari:2008rk} \\
\hline
\end{tabular}
\caption{Table of sources used in this paper}\label{table:sources}
\end{center}
\end{table}

\subsection{Analysis}
A similar analysis was done by the \textit{Fermi} collaboration \cite{Ackermann:2012sza} for sources in the high energy (HE) regime and around a redshift, $z\sim1.0$. Since \textit{Fermi} has an energy range $\sim$ 100s MeV to $\sim$ 100s GeV, they are able to assume the low energy spectrum of the source, where little aborption is expected, is unabsorbed. They can then construct an intrinsic spectrum under the assumption that the spectral slope holds to energies $\sim$ 100s GeV in order to measure the energy dependent attenuation. Since we are concerned with VHE sources which begin at the upper end of Fermi's reach, the entire VHE spectrum is likely absorbed and we are unable to make such assumptions. Instead, we fit a model for the attenuation of the form 

\begin{equation}
F_{theory}=N(\Gamma_{i}, b_{i})E^{-\Gamma_i}e^{-b_{i}\tau_{theory}}, 
\end{equation}
to the observed spectrum of each source.
We let the spectral index of the source, $\Gamma$, vary (1.5 - 4) as it is ultimately a nuisance parameter for our purposes. The factor $N$ is an overall normalization factor that encompasses the total flux measured from a source while also normalizing the spectrum for a given combination of $\Gamma$ and $b$.  $N$ is found by integrating the spectrum, for a given $\Gamma$ and $b$,  where $b$ parameterizes the EBL models by scaling the predicted optical depth for a given model.   For the predicted optical depths we use a representative suite of models
from Franceschini \cite{Franceschini:2008tp}, Dominguez \cite{Dominguez:2010bv}, and Gilmore \cite{Gilmore:2009zb}. A value of $b=1$ indicates the sources completely agree with the model, values $b > 1$ indicates more attenuation is required to fit the sources than that predicted by a given model, while $b <1$ indicates the sources prefer less attenuation.   We find the best $b$ by doing a global fit and then a marginalized likelihood, integrating out the spectral indicies. 
 
\begin{figure}
\centering
\includegraphics[width=0.65\textwidth]{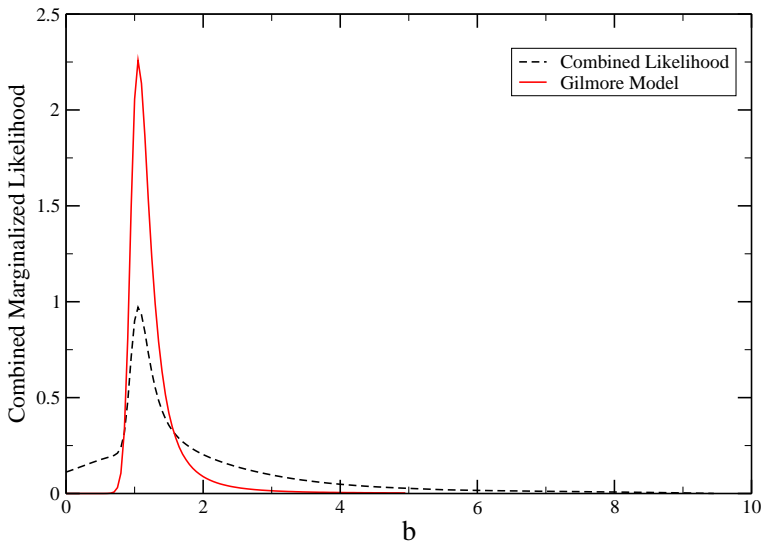}
\caption{The marginalized likelihood for the three models combined and for the Gilmore model, alone. }
\label{fig:likeli}
\end{figure}

\begin{figure}[h!]
 \centering
 \begin{subfigure}{0.9\textwidth}
   \centering
   \includegraphics[width=0.85\textwidth]{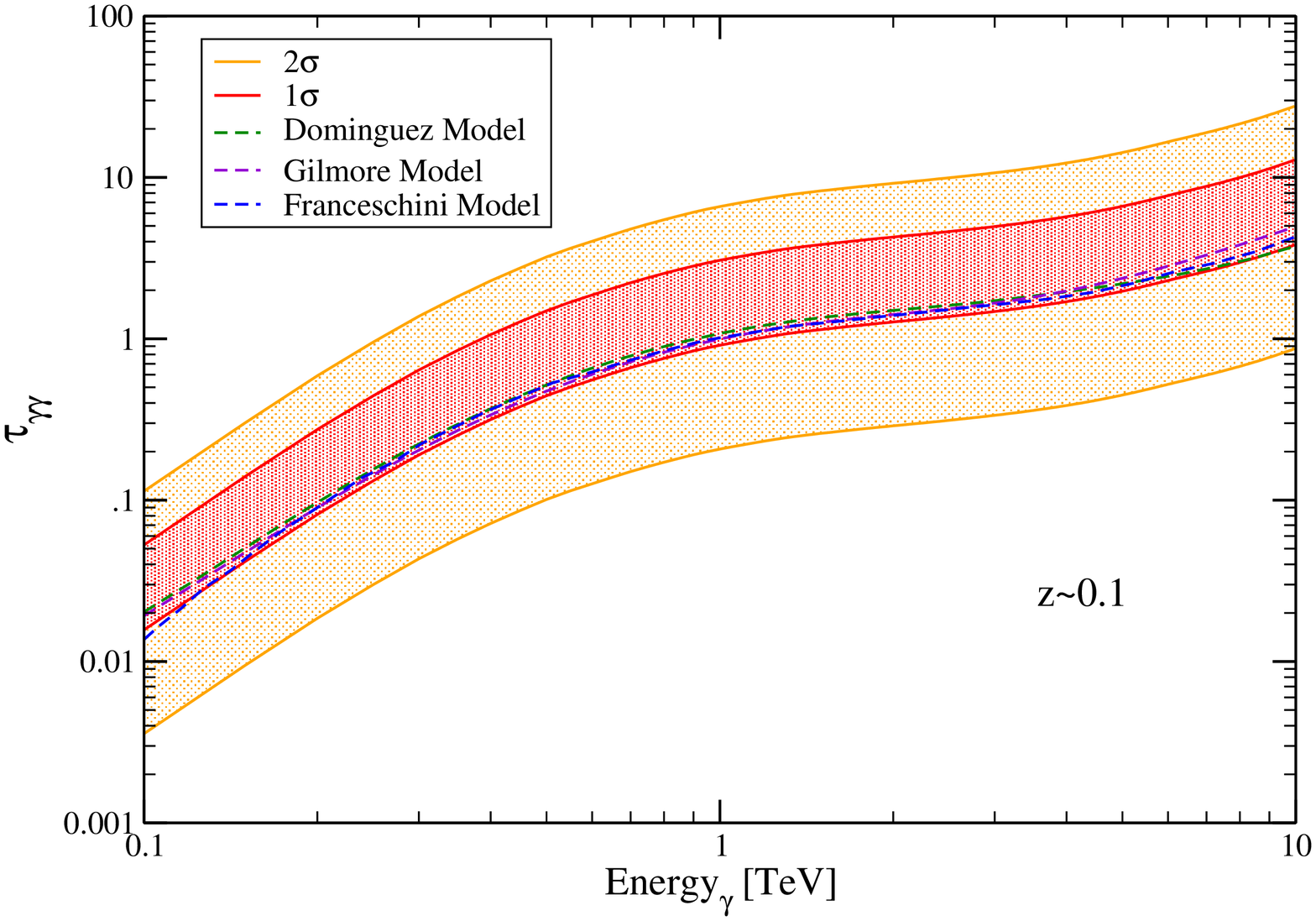}

   \par\addvspace{1.4cm}
   \includegraphics[width=0.85\textwidth]{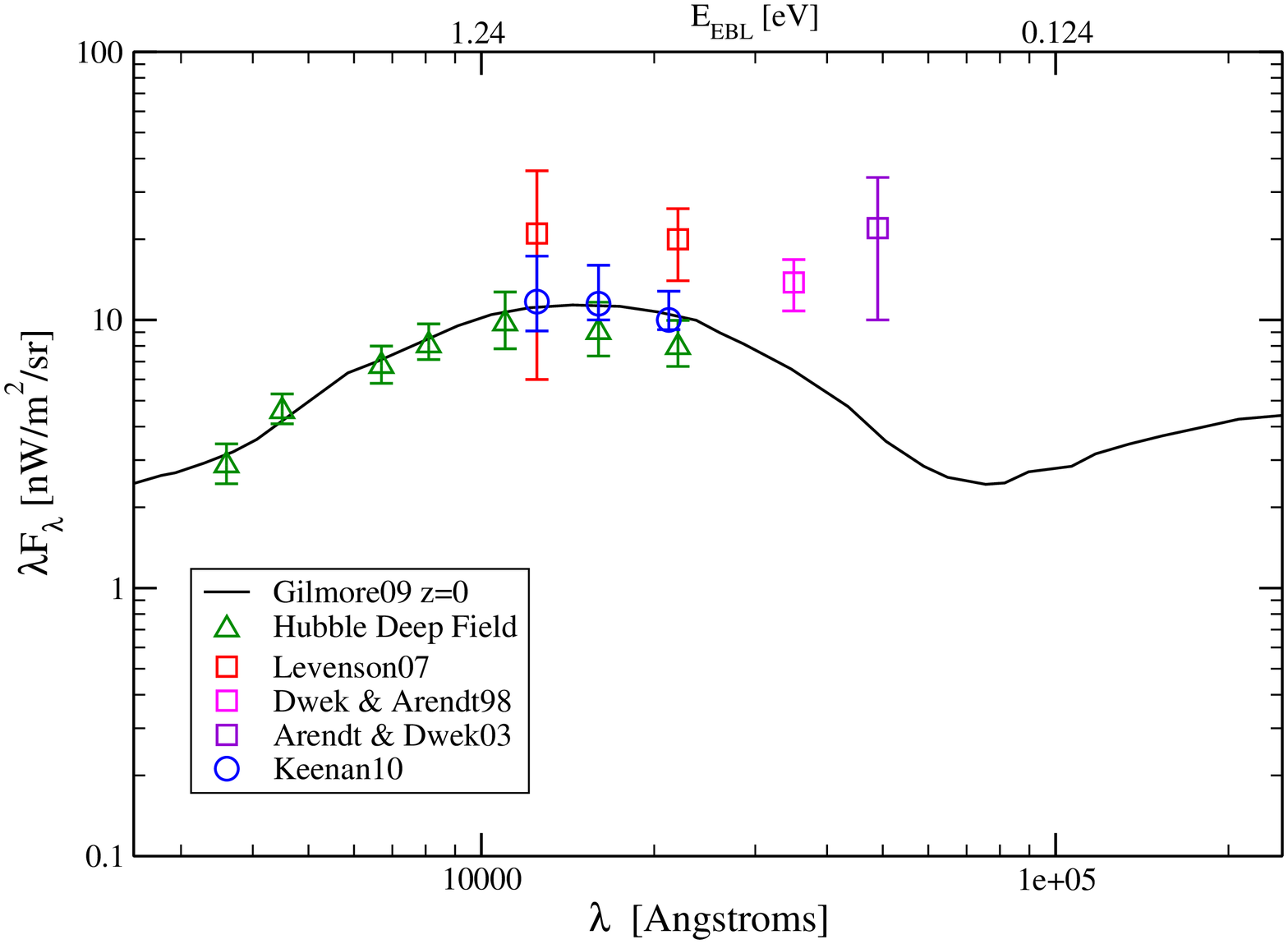}
 \end{subfigure}
\caption{\textit{Top:} The optical depth $\tau_{\gamma \gamma}$ as a function of gamma-ray energy at $z \approx 0.1$  for all three models combined. \textit{Bottom:} Gilmore09's EBL model aligned so that the EBL energy range corresponds to the threshold of pair-production for the given gamma-ray energy in the top panel. Integrated galactic light from the Hubble deep field constraints \cite{Madau:1999yh} are shown as the green triangles, and the Keenan10 points are from Subaru \cite{Keenan:2011ih}. Absolute measurements from COBE-DIRBE are shown as squares, \textit{the different colors correspond to different references} \cite{Levenson:2007is, Dwek:1998sr, Arendt:2002mk}.}
\label{fig:tau}
\end{figure}

In Fig. \ref{fig:likeli} we show the combined marginalized likelihood of $b$ from the three models.  All the models from our suite adequately describe the data.  The maximum best fit $b$ values, and 1$\sigma$ bounds, are $b=1.31^{+2.42}_{-0.60}$, $b=1.21^{+2.44}_{-0.55}$, and $b=1.05^{+0.41}_{-0.07}$ for the Franceschini, Dominguez, and Gilmore models, respectively. The 1$\sigma$ and 2$\sigma$ ranges were found by integrating the likelihood curve and finding the middle 68$\%$ and 95$\%$ containment ranges, respectively. The best fit value for the combined likelihood, and 1$\sigma$ bounds, is $b=1.05^{+1.92}_{-0.17}$.  We combined the models, giving them equal weight, and multiplied the $b$ values by the theoretical $\tau$ curves to calculate the best fit $\tau$'s as presented in the top panel of Fig. \ref{fig:tau}.

The statistical significance of a given model is complicated by the non-gaussian nature of the likelihood.  We can construct a test statistic (TS) from the likelihood ratio test, given by ${\rm TS}=2\log\left[\mathcal{L}(b=b_{max})/\mathcal{L}(b=0)\right]$. 
 If we approximate the likelihood curve as being gaussian the probability distribution of the TS follows a chi-squared distribution. If we make this approximation for our Gilmore fit, which has the tightest likelihood curve, we can exclude the null hypothesis ($b=0$) at the 7$\sigma$ level (TS=52).

\begin{figure}
\centering
\includegraphics[width=0.65\textwidth]{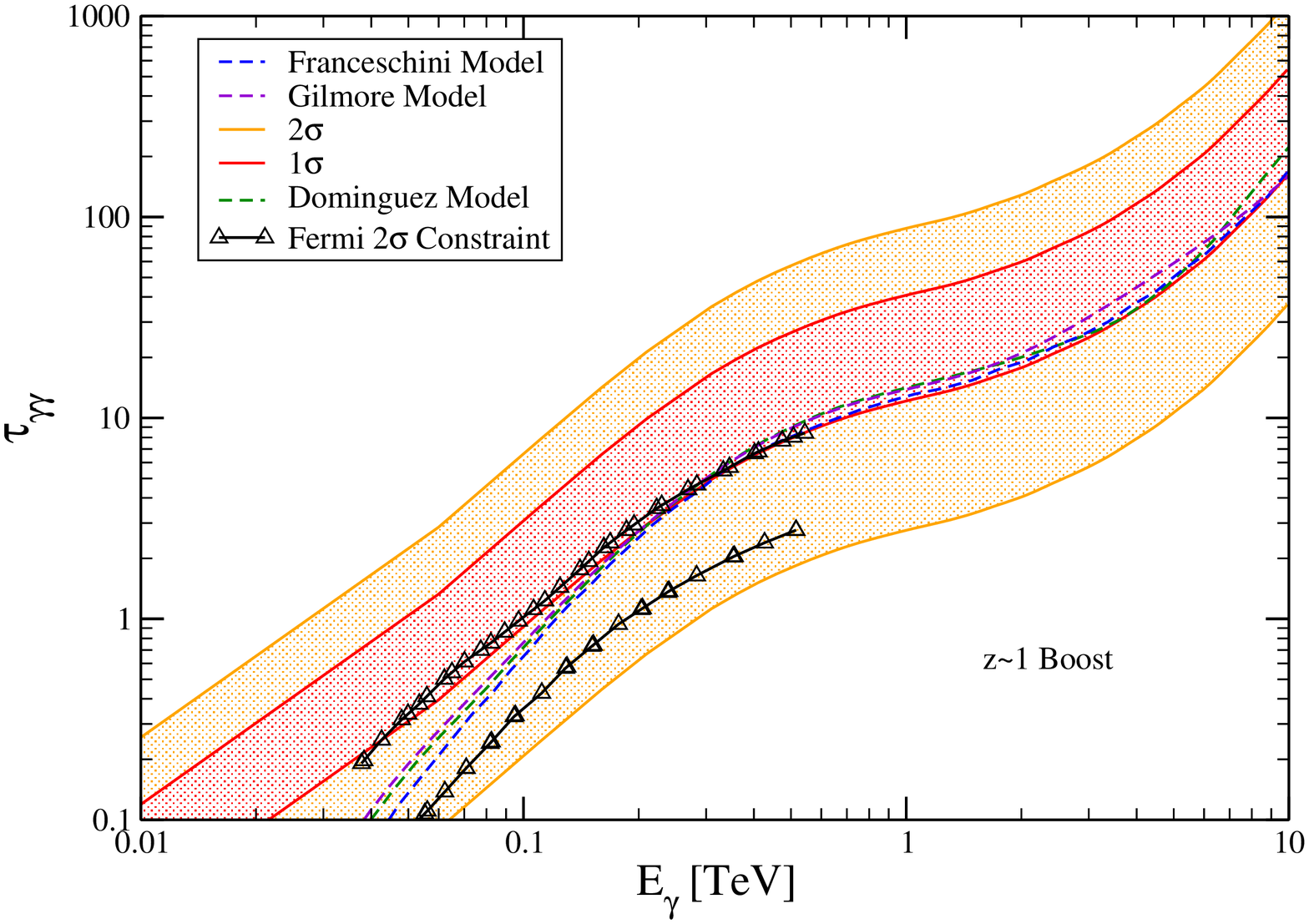}
\caption{The optical depth as a function of gamma-ray energy fit as in Fig.\ref{fig:tau}, boosted to $z \approx 1.0$. The \textit{Fermi} results \cite{Ackermann:2012sza} are shown by black lines and triangles. }
\label{fig:boost}
\end{figure}

Using our suite of EBL models, the optical depth at $z \sim 0.1$ presented here can predict the optical depth at $z \sim 1$ so that our results can be directly compared to those of the \textit{Fermi} collaboration \cite{Ackermann:2012sza}. We did this by taking our fit $b$ values from $z=0.1$ and applied them to the same EBL models at $z=1$.  We show the optical depth predicted at $z \sim 1$ using our TeV point sources along with the \textit{Fermi} result in Fig. \ref{fig:boost}.  The two measurements show good agreement, and combined in this way, measure the energy density of the EBL from 0.05 eV -- 50 eV.

\section{Conclusion and Discussion}
We have shown how to combine a set of TeV blazar measurements in order to indirectly probe the optical depth of the Universe to VHE gamma-rays. We assumed an EBL spectral shape from three frequently used models. The TeV sources require attenuation which is consistent with the models predictions for the EBL. The primary source of the EBL in this energy range is thought to be starlight  and it appears that the models provide a good description, as measured by TeV blazars.  Specifically, the star formation history of the Universe as assumed/calculated in our suite of models is consistent with the attenuation of TeV blazars.
Our data set was intentionally chosen  to cluster at redshift $\sim0.1$. When we boost our constrained models to a redshift $z\sim 1$, we show that the models, as constrained by TeV sources, are in good agreement with the results obtained by \textit{Fermi}.  In particular, they both agree for  gamma-rays which interact with EBL photons with energies of a few eV.

Indirect constraints of the EBL using attenuation of TeV sources have been previously obtained.  A group of  studies \cite{Georganopoulos:2010iv, Yang:2010ur} extrapolate the \textit{Fermi} measured spectrum of a source out to TeV energies, resulting in a single power law source spectrum. They assume this extrapolation to be an upper limit to the intrinsic TeV spectrum thus giving them a strict upper limit to the EBL absorption.  Another technique is to examine the nature of the connection between the HE spectrum of a source to its VHE spectrum \cite{Senturk:2013pa, Fermi-LAT:2010gba, Dwek:2012nb}, namely, whether or not a single power law is justified or if a broken power law model should be invoked. Other attempts, including work by \cite{Mazin:2007pn}, use a spline method to adjust the shape of the EBL such that when they deabsorb a TeV blazar, the resulting intrinsic spectrum is physically feasible.  Work by \cite{Dominguez:2013lfa} measures the cosmic $\gamma$-ray horizon, independent of any EBL model, using 15 blazars and predicting the intrinsic VHE blazar spectrum assuming a synchrotron self-Compton model. 
Recently, the H.E.S.S. collaboration \cite{Abramowski:2012ry} used a similar analysis to that presented here for a set of 7 blazars, having one member in common with our data set, split into groups by redshift ($z \sim .05, .12,$ and .17).   They find constraints on the energy density of the EBL from $\sim$ 0.1 eV - 5 eV which are consistent with our analysis for the optical depth.  
In all cases, the constraints obtained on the EBL are consistent with those presented in this work.   Our analysis, along with those of \textit{Fermi} and H.E.S.S., indicate absorption by the EBL at $6 - 7 \sigma$.
The analysis presented here is somewhat less dependent on the exact spectrum of the source.  In addition, it presents a measurement of the optical depth for TeV gamma-rays due the EBL from a combined set of sources and shows that the \textit{Fermi} results, taken at lower energies and higher redshift, are perfectly consistent with TeV measurements for the same set of models.

\section{Acknowledgements}
We would like to thank James Stapleton for helpful discussions on the likelihood analysis.

\bibliographystyle{plain}
\bibliography{EBL_bib_long}

\end{document}